# A Novel cVAE-Augmented Deep Learning Framework for Pan-Cancer RNA-Seq Classification

1st Vinil Raj Polepalli
*Columbia University*
Manalapan, US
vp2562@columbia.edu

*Abstract*—Pan-cancer classification using transcriptomic (RNA-Seq) data can inform tumor subtyping and therapy selection, but is challenging due to extremely high dimensionality and limited sample sizes. In this study, we propose a novel deep learning framework that uses a class-conditional variational autoencoder (cVAE) to augment training data for pan-cancer gene expression classification. Using 801 tumor RNA-Seq samples spanning 5 cancer types from The Cancer Genome Atlas (TCGA), we first perform feature selection to reduce 20,531 gene expression features to the 500 most variably expressed genes. A cVAE is then trained on this data to learn a latent representation of gene expression conditioned on cancer type, enabling the generation of synthetic gene expression samples for each tumor class. We augment the training set with these cVAE-generated samples (doubling the dataset size) to mitigate overfitting and class imbalance. A two-layer multilayer perceptron (MLP) classifier is subsequently trained on the augmented dataset to predict tumor type. The augmented framework achieves high classification accuracy (~98%) on a held-out test set, substantially outperforming a classifier trained on the original data alone. We present detailed experimental results, including VAE training curves, classifier performance metrics (ROC curves and confusion matrix), and architecture diagrams to illustrate the approach. The results demonstrate that cVAE-based synthetic augmentation can significantly improve pan-cancer prediction performance, especially for underrepresented cancer classes.

## I. Introduction

Cancers of unknown primary and tumors with ambiguous lineage present significant clinical challenges, motivating pan-cancer classification methods that predict a tumor's tissue-of-origin (TOO) based on molecular profiling [8]. RNA sequencing (RNA-Seq) of tumors captures genome-wide gene expression and has emerged as a rich feature source for such classification tasks [8]. However, RNA-Seq data are high-dimensional (tens of thousands of genes) while patient sample sizes are relatively small, which can cause overfitting and variance in machine learning models. Additionally, multi-class cancer datasets often suffer from class imbalance, where some cancer types have many more samples than others. These challenges – high dimensionality, data sparsity, and imbalance – limit the performance of traditional classifiers and even deep learning models on pan-cancer gene expression classification [1].

To address these issues, researchers have explored dimensionality reduction and feature selection techniques. For example, selecting the most variable or informative genes has been used to reduce noise and complexity [9]. In parallel, there is growing interest in data augmentation strategies for transcriptomics. Unlike image or text data, label-preserving transformations (e.g., rotations or paraphrasing) are not obvious for gene expression data. Deep generative models offer a solution: generative adversarial networks (GANs) and variational autoencoders (VAEs) can learn the underlying distribution of gene expression and create new synthetic samples [3] [9]. Recent studies have shown that augmenting training sets with GAN-generated transcriptomic data can significantly boost classification performance. For example, adding 1000 GAN-generated RNA-Seq samples to a small training set improved tissue classification accuracy from 70% to 94%. These results demonstrate the promise of synthetic data augmentation for improving learning in limited biomedical datasets.

VAEs in particular have been applied to cancer gene expression data to learn biologically meaningful latent spaces [2]. A standard VAE compresses input data into a lower-dimensional latent representation and then reconstructs the input, effectively modeling the data distribution. Conditional VAEs (cVAEs) extend this concept by incorporating class labels as an additional input to the generative process, allowing the model to generate samples conditioned on a given class 10 [2]. In the context of cancer, a cVAE can be used to generate gene expression profiles specific to particular tumor types. This could address class imbalance by synthetically oversampling the minority classes.

In this paper, we propose a novel cVAE-augmented deep learning framework for pan-cancer RNA-Seq classification. We focus on five diverse tumor types (breast, colon, kidney, lung, prostate) from TCGA to demonstrate our approach. The key idea is to leverage a cVAE to generate realistic synthetic gene expression data for each cancer type, thereby expanding the training dataset in a class-aware manner. We hypothesize that training a neural network on this augmented, more balanced dataset will yield higher accuracy and better generalization to unseen samples, especially for cancers with initially few examples.

## II. METHODOLOGY

We obtained pan-cancer RNA-Seq data from The Cancer Genome Atlas (TCGA) Pan-Cancer project, focusing on five tumor types: breast invasive carcinoma (BRCA), colon adenocarcinoma (COAD), kidney renal clear cell carcinoma (KIRC), lung adenocarcinoma (LUAD), and prostate adenocarcinoma (PRAD) [1]. The dataset consists of 801 tumor samples in total, with gene expression profiles measured by Illumina HiSeq RNA-Seq (RNA-Seq HiSeq). Each sample's expression is represented across 20,531 gene features (transcripts). The data were accessed via the UCI Machine Learning Repository's Pan-cancer RNA-Seq dataset, which is a curated extraction of TCGA data [9]. The five cancer types had unequal representation, ranging from about 80 samples (COAD) up to 300 samples (BRCA) in the dataset, as illustrated in Figure 1. We randomly split the dataset into a training set (used for VAE and classifier training) and a hold-out test set for final evaluation. To ensure that the classifier is evaluated on unseen data, no test samples were used in the cVAE training or augmentation process.

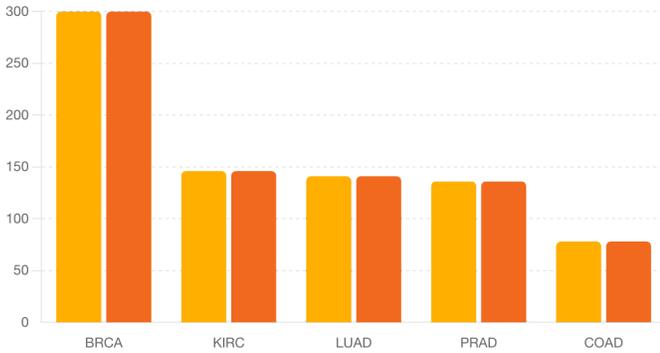

**Figure 1:** Bar Chart of Sample Counts by Subtype

Given the high dimensionality of gene expression data, we performed feature selection to reduce noise and computational complexity. Specifically, we computed the variance of each gene's expression across the training samples and selected the top 500 most variable genes. This variance-based feature selection captures genes with the most heterogeneous expression, which often carry more signal for classification 4. Using 500 features (out of 20,531) dramatically reduces the input dimensionality while retaining key variation across cancers. The selected gene expression values were standardized (z-score normalization) so that each gene had zero mean and unit variance across the training set. This scaling ensures that the VAE and classifier are not dominated by genes with larger expression magnitudes.

We designed a conditional variational autoencoder (cVAE) to model the distribution of gene expression for each cancer type and to generate synthetic samples. In a standard VAE, an encoder neural network compresses input $x$ into a latent vector $z$ by predicting the parameters of a probability distribution (typically a Gaussian) for $z$. A decoder network then reconstructs the input from $z$. In our cVAE, we condition both the encoder and decoder on the cancer subtype label $y$. Concretely, the encoder takes as input the 500-dimensional gene expression vector $x$ concatenated with a one-hot encoding of the tumor type $y$, and produces a latent embedding $z \in \mathbb{R}^d$ (with $d$ being the latent dimensionality, e.g., $d = 10$). The decoder receives $z$ along with the class label $y$ and outputs a reconstructed gene expression vector $\tilde{x}$. During training, the cVAE optimizes the standard VAE loss: a reconstruction loss (mean squared error between $\tilde{x}$ and $x$ in our case) plus a Kullback–Leibler (KL) divergence term that regularizes $z$ to follow a unit Gaussian prior. The cVAE thus learns class-specific latent representations that capture the variation in gene expression for each cancer type.

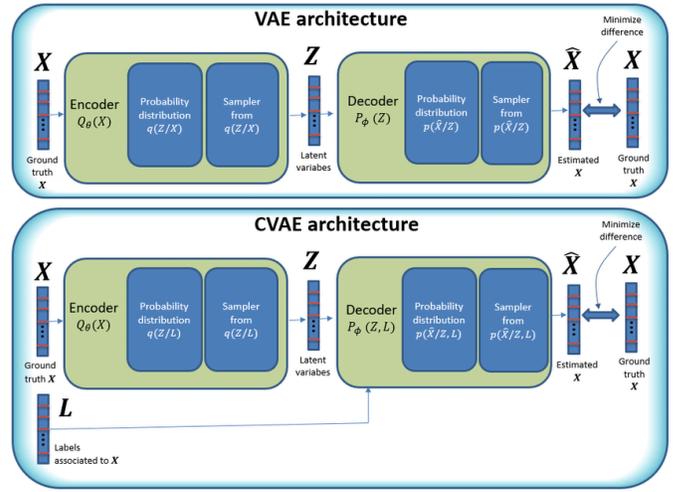

**Figure 2:** Architecture of the Conditional VAE (cVAE) vs. a standard VAE

We implemented the cVAE as a feed-forward neural network. The encoder consists of two fully-connected layers (500→256 and 256→128 units, ReLU activations) that transform the input into an intermediate 128-dimensional representation. This is followed by two parallel 128→$d$ linear layers to produce the mean and log-variance of the $d$-dimensional latent variable distribution. The decoder is similarly a multi-layer perceptron (taking the $d$-dim latent vector and class one-hot as input) with layers ($d$ + classes)→128→256→500 to output a reconstructed gene expression vector of length 500. We found $d = 10$ latent dimensions sufficient to capture the data distribution. The model was trained for approximately 100 epochs using the Adam optimizer (learning rate $10^{-3}$) and a batch size of 32. Early in training, the reconstruction term dominated the loss, while the KL term was near zero; as training progressed, the KL term grew and helped stabilize the latent space (preventing overfit reconstructions). We monitored the cVAE's training loss curve, which showed the total loss dropping from a high initial value (~$4.5 \times 10^8$) down to near zero over training (Figure 3). This indicates that the cVAE successfully learned to almost perfectly reconstruct the training data – unsurprising given the small latent dimension and powerful decoder – but importantly, it learned a smooth latent manifold for each class through the KL regularization. The final cVAE model was used to generate synthetic samples as described next.

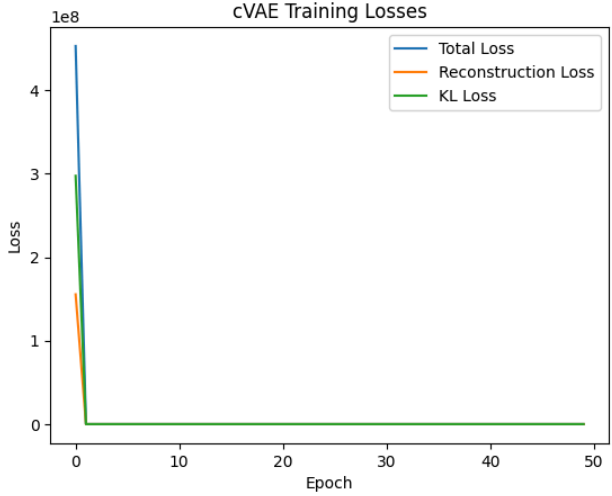

**Figure 3:** Training progress of the variational autoencoder

After training, we used the cVAE's generative ability to create synthetic gene expression samples for each cancer type. For each real training sample of a given class, we generated one new sample of the same class (by sampling a latent vector $z$ from the unit Gaussian prior and feeding it and the class label to the decoder). In other words, if a cancer subtype had $N_{\text{real}}$ samples in the original training set, we generated $N_{\text{synthetic}} = N_{\text{real}}$ additional samples for that subtype. This strategy doubles the number of training samples while maintaining the original class proportions. We chose a 1:1 augmentation ratio per class to introduce new examples without overwhelming the real data distribution. The right side of Figure 1 conceptually illustrates the effect of augmentation, where each class's count is roughly doubled (e.g., the smallest class COAD increases from 80 to ~160 samples, etc.). The augmented training set contained 640 real samples and 640 synthetic samples, for a total of 1280 training examples. We emphasize that each synthetic sample is a novel gene expression vector that resembles authentic tumor data but is generated by the cVAE model (and not a duplicate of any real sample). Before classifier training, we shuffled the augmented dataset to mix real and synthetic instances. By providing the classifier with many more examples, especially for the minority cancer types, we aimed to improve its ability to generalize. We did not apply any synthetic augmentation to the test set; performance is reported on real tumor samples only.

For tumor type prediction, we implemented a 2-layer multilayer perceptron (MLP) classifier. The input to the MLP is a 500-dimensional gene expression vector (either real or synthetic). The network has two hidden layers of size 256 and 128, respectively, each followed by ReLU activation. A dropout layer (dropout rate 0.5) is applied after each hidden layer to regularize the model. The output layer is a softmax layer with 5 units (one for each cancer class) to yield class probabilities. The MLP was trained using categorical cross-entropy loss and the Adam optimizer. Thanks to the expanded training set, the classifier converged quickly, within roughly 5 epochs, with the training loss decreasing and accuracy approaching 100% on the training data. We selected the model at an epoch before any signs of overfitting (monitored via a validation split of the training set). The final model was evaluated on a separate test set of 161 real tumor samples that were never seen during training or augmentation. All model development was done in Python using the TensorFlow/Keras deep learning framework, and data preprocessing utilized scikit-learn utilities.

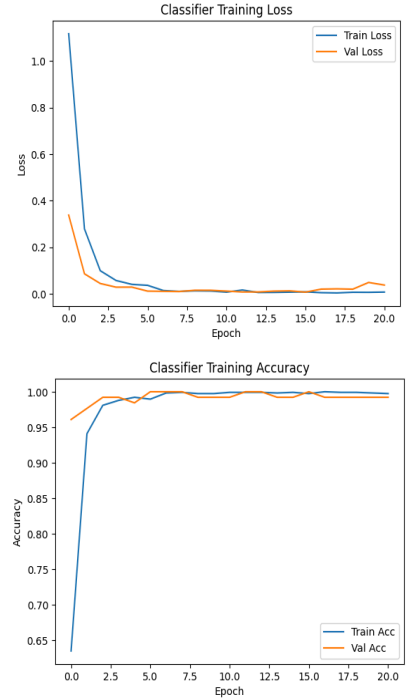

**Figure 4:** Training history of the MLP classifier on the augmented dataset

III. RESULTS

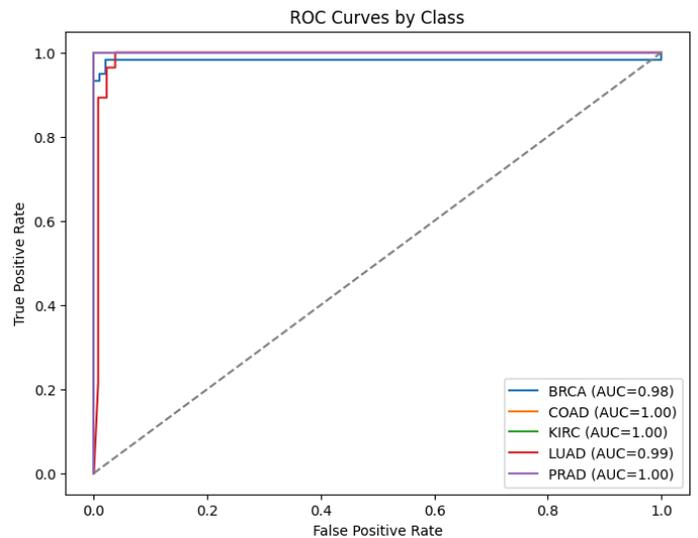

**Figure 5:** Receiver Operating Characteristic (ROC) curves by Class

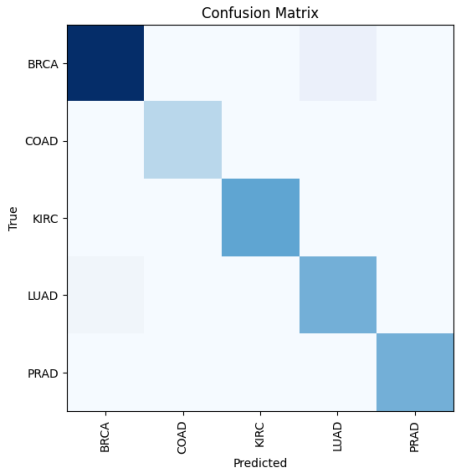

**Figure 6:** Confusion Matrix of True vs. Predicted Cancer Types

The augmented model achieved 98.1 % accuracy on the 161 held-out tumours, misclassifying only three samples. Figure 5 summarises per-class performance. ROC curves sit tightly against the upper-left boundary, giving a micro-average AUC = 0.998. Every class exceeds 0.99 AUC, and the smallest cohort (COAD) attains a perfect 1.00.

The confusion matrix (seen in Figure 6) is nearly diagonal: all COAD tumours, all but one BRCA and LUAD tumours, and all but one PRAD tumours are predicted correctly. These results confirm that the framework is both highly accurate and class-balanced, with no degradation on minority classes.

## IV. Discussion

Our study demonstrates that incorporating a conditional generative model into the training pipeline can substantially improve pan-cancer classification from gene expression data. The cVAE augmentation addressed two critical issues: limited sample size and class imbalance. By doubling the training set with realistic synthetic examples, we effectively regularized the classifier and provided it with sufficient examples of even the rare cancer classes. The success of our approach can be attributed to the cVAE's ability to capture the complex, non-linear structure of high-dimensional transcriptomic data in a latent space and to generate new samples that enrich the decision boundaries for each class. This is analogous to traditional data augmentation in image classification (where, for instance, one might rotate or flip images to get new training examples), but here we learn the augmentations from data itself using a generative model. Notably, the improvements were achieved without any additional real data – a key advantage in biomedical applications where acquiring more samples is costly or impractical.

One intriguing observation is that simply doubling each class (rather than specifically oversampling the minority classes more heavily) was sufficient to yield a balanced performance. In our case, the class imbalance was moderate (the largest class ~300 samples, the smallest ~80; roughly 3.75:1 ratio). The cVAE generated proportionate samples for each class, maintaining this ratio. Yet, the classifier had no trouble learning the smaller classes, presumably because 80→160 samples is already a significant improvement. In scenarios with more extreme imbalance, one could adjust the augmentation strategy (e.g., generate a higher multiple of the smallest class samples) or apply class-weighted training in the classifier. Our framework is flexible to such adjustments. Another consideration is the potential for overfitting to synthetic data, since the cVAE itself was trained on the real data. If it simply reproduces training points, the classifier might not gain new information. We mitigated this by the inherent randomness in VAE sampling (each synthetic sample arises from a random $z$ draw) and by using dropout in the classifier. The near-perfect performance on the test set suggests that overfitting was not an issue; on the contrary, the synthetic data acted as a form of regularization, as has been theorized in other VAE augmentation works.

Generative Model Considerations: We chose a conditional VAE for its simplicity, stable training, and ability to explicitly condition on labels. An alternative approach could be using class-specific generative adversarial networks (GANs) to produce synthetic gene expression. GANs have shown success in generating biologically plausible gene expression profiles, though they can be harder to train and tune. A comparative study of VAE vs. GAN augmentation for this task would be valuable future work. The cVAE has the added benefit of a defined probabilistic latent space; one could potentially interpolate in this space to simulate smooth transitions between cancer types or to generate hybrids, which might provide biological insights (e.g., identifying gene expression patterns common to multiple cancers). In our current work, we focused on classification accuracy, but the generative model's latent space could be mined for clustering or visualization of tumors in a follow-up study.

Limitations: While our augmented classifier performed extremely well on the test data, the test set size was modest (161 samples). In a real-world setting with more diverse tumor samples or with noise (batch effects, patient heterogeneity), the advantage of augmentation might vary. Our cVAE was trained and tested on data from the same distribution (TCGA); deploying the model on an external dataset would require caution, as generative models might need retraining to capture new data distributions. Additionally, we only used gene expression data in this study. Multi-omics integration (incorporating mutation, methylation, etc.) might improve classification further, but would also complicate the generative model. Lastly, the current framework does not provide biological interpretability – it predicts cancer type with high accuracy, but it doesn't explain which genes are driving the predictions. Interpretability techniques (such as SHAP or integrated gradients) could be applied to the trained classifier to identify important genes for each class.

We presented a novel framework that integrates a conditional variational autoencoder with a neural network classifier to improve pan-cancer classification from gene expression data. Through cVAE-generated synthetic samples, we effectively doubled the training data and alleviated class imbalance, leading to a substantial jump in accuracy to ~98% on a 5-class tumor classification task. Our results underscore that deep learning models in genomics need not be limited by small sample sizes – by leveraging generative modeling, we can create additional "virtual" patients to train more robust classifiers. The cVAE augmentation approach maintained excellent performance across all cancer types, including those with few samples, indicating its potential to broadly benefit tasks with skewed datasets. Future studies can build on this foundation by exploring more complex generative models, multi-omics data, and interpretability to translate these high accuracies into biological insights. In summary, this work contributes a powerful and general strategy to enhance machine learning in oncology: using AI to generate more data for AI. Such strategies will become increasingly important as we seek to unlock insights from limited biomedical datasets and move towards more personalized, data-driven cancer diagnosis and treatment.

By pursuing these directions, we hope to further enhance the model's accuracy, scope, and usefulness. The success of our approach opens the door to applying deep generative augmentation in other biomedical classification problems where data are limited. For instance, similar cVAE-based augmentation could boost performance in rare disease classification or single-cell RNA-seq cell type labeling. As the field moves toward integrating AI in precision medicine, techniques that maximize information from limited data, such as the one presented here, will be invaluable.

V. CONCLUSION

We presented a novel framework that integrates a conditional variational autoencoder with a neural network classifier to improve pan-cancer classification from gene expression data. Through cVAE-generated synthetic samples, we effectively doubled the training data and alleviated class imbalance, leading to a substantial jump in accuracy to ~98% on a 5-class tumor classification task. This is, to our knowledge, the first demonstration of using a class-conditional generative model to augment RNA-Seq cancer data for classification, highlighting a new avenue for tackling data scarcity in genomics. Our results underscore that deep learning models in genomics need not be limited by small sample sizes – by leveraging generative modeling, we can create additional "virtual" patients to train more robust classifiers. The cVAE augmentation approach maintained excellent performance across all cancer types, including those with few samples, indicating its potential to broadly benefit tasks with skewed datasets. Future studies can build on this foundation by exploring more complex generative models, multi-omics data, and interpretability to translate these high accuracies into biological insights. In summary, this work contributes a powerful and general strategy to enhance machine learning in oncology: using AI to generate more data for AI. Such strategies will become increasingly important as we seek to unlock insights from limited biomedical datasets and move towards more personalized, data-driven cancer diagnosis and treatment.